\documentclass{mem}
\usepackage{newtxtext,newtxmath}

\usepackage{mathtools}
\usepackage{physics}

\usepackage{natbib}
\usepackage{balance}
\usepackage{flushend}
\usepackage{graphicx}
\usepackage{bm}

\usepackage{hyperref}
\hypersetup{colorlinks = true,
            allcolors = blue,}
\idline{0}{1}
\begin{document}

\title{
Exploring the impact of electromagnetic dissipation on ultra-relativistic plasma outflows
}

\author{
Argyrios Loules \and Nektarios Vlahakis
          }

\institute{Section of Astrophysics, Astronomy and Mechanics, Department of Physics, National and Kapodistrian University of Athens, University Campus, Zografos, GR-15784}

\authorrunning{Loules \& Vlahakis}

\titlerunning{EM dissipation in ultra-relativistic plasma outflows}


\abstract{
Ultra-relativistic plasma outflows are intrinsically connected with gamma-ray bursts. Over the years, a large number of analytical and numerical works has been devoted to understanding the intricacies of their complex dynamics, with most of these past studies performed in the ideal MHD regime. We propose a self-similar formalism, based on the expansion of the equations of resistive relativistic magnetohydrodynamics, for the description of these outflows in the vicinity of their symmetry axis and present semi-analytical solutions describing strongly relativistic jets in both the ideal and resistive MHD regimes. Our solutions provide a clear picture of the impact of electromagnetic dissipation on the acceleration and collimation mechanisms which determine the kinetic and morphological characteristics of these relativistic outflows. The resistive MHD solutions are compared to their ideal MHD counterparts, revealing the key differences between the two regimes. Our comparative analysis sheds light on the possible role of electromagnetic dissipation in shaping the dynamics of the ultra-relativistic outflows associated with gamma-ray bursts.

}
\maketitle{}

\section{Introduction}

Gamma-ray bursts (GRBs) are intrinsically connected to ultra-relativistic outflows of plasma \citep{piran2004}, launched either following the collapse of a massive star (long GRBs) \citep{macfadyen1999} or after binary mergers \citep{eichler1989}. The central role that ultra-relativistic plasma outflows possess in shaping the emission characteristics of these extreme astrophysical events \citep{pawan2015} has led to their extensive analytical \citep{vlahakis2003, globus2014} and numerical \citep{komissarov2009, gottlieb2020} modeling in the context of ideal relativistic magnetohydrodynamics (MHD). Recently, the dynamics of resistive relativistic plasma outflows associated with GRBs were investigated by \cite{mattia2024}, who utilized resistive relativistic MHD simulations to study the effects of magnetic dissipation on their properties.

In this work, we present exact solutions which describe ultra-relativistic resistive jets in the vicinity of their symmetry axis. The solutions were obtained by use of a self-similar formalism based on the angular expansion of the equations of steady state and axisymmetric general-relativistic resistive MHD (GR-RMHD). Our results showcase the impact of ohmic or electromagnetic (EM) dissipation on the acceleration and collimation profiles of ultra-relativistic plasma outflows.

\section{Self-similar modeling and results}

Following the standard $3+1$ approach, we write the equations of motion of general relativistic resistive GR-RMHD in the static observer frame for a Schwarzschild spacetime:

\begin{equation}\label{continuity}
    \bm{\nabla}\cdot(h_{t}\varGamma\rho\bm{v}) = 0\, ,
\end{equation}

\begin{align}
    \varGamma\rho\bm{v}\cdot\bm{\nabla}(\xi\rho\bm{v}) = &-\bm{\nabla}P + \dfrac{J^{0}\bm{E} + \bm{J}\cross\bm{B}}{c} \notag \\
    &\quad - \varGamma^{2}\rho\xi c^{2}\bm{\nabla}\ln{h_{t}}\, .
\end{align}

\begin{align}
    \varGamma\rho c^{2}\bm{v}\cdot\bm{\nabla}\xi - \varGamma\bm{v}\cdot\bm{\nabla}P = &\varGamma\bm{J}\cdot\left(\bm{E} + \dfrac{\bm{v}\cross\bm{B}}{c}\right) \notag \\ & \quad - \varGamma J^{0}\dfrac{\bm{E}\cdot\bm{v}}{c}\, ,
\end{align}
supplemented by Maxwell's equations
\begin{equation}
    \bm{\nabla}\cdot\bm{B} = 0\, ,
\end{equation}

\begin{equation}
    \bm{\nabla}\cdot\bm{E} = \dfrac{4\pi}{c}J^{0}\, ,
\end{equation}

\begin{equation}
    \bm{\nabla}\cross(h_{t}\bm{E}) = 0\, ,
\end{equation}

\begin{equation}\label{ampere}
    \bm{\nabla}\cross(h_{t}\bm{B}) = \dfrac{4\pi h_{t}}{c}\bm{J},
\end{equation}
where $\bm{v}$, $\bm{B}$, $\bm{E}$, $J^{0}/c$, and $\bm{J}$ are the plasma flow three-velocity, magnetic and electric fields, charge density, and current density, as measured by the static observer. $\rho$ and $P$ are the plasma's rest mass density and thermal pressure, $\varGamma = \left(1 - v^{2}/c^{2} \right)^{-1/2}$ is the flow Lorentz factor, and $h_{t} = \sqrt{1 - r_{S}/r} = \sqrt{1 - 2GM/(c^{2}r)}$ is the lapse function. Closure of the system of equations is achieved by the inclusion of Ohm's law \citep{bucciantini2013}
\begin{equation}\label{ohm}
    \bm{J} = J^{0}\dfrac{\bm{v}}{c} + \dfrac{1}{4\pi\eta}\left(\bm{E} + \dfrac{\bm{v}\cross\bm{B}}{c}\right)\, ,
\end{equation}
with $\eta$ the plasma's electrical resistivity, and of a polytropic, variable adiabatic index equation of state \citep{ryu2006}
\begin{equation}
    \xi = 2\dfrac{6\Theta^{2} + 4\Theta + 1}{3\Theta + 2}\, ,
\end{equation}
where $\Theta = P/(\rho c^{2})$ is the plasma's dimensionless temperature.

Eqs. \ref{continuity}-\ref{ohm} constitute a system of non-linear partial differential equations in the radial coordinate, $r$, and polar angle, $\theta$. In order to achieve the separation of variables, we assume profiles for the magnetic flux, mass flux, and electric potential functions, as well as for the rest mass density, thermal pressure, toroidal magnetic field and velocity components, and finally for the resistivity. These profiles consist of an unknown function of $r$ and a prescribed dependence on $\theta$. We then expand Eqs. \ref{continuity}-\ref{ohm} with respect to the polar angle, keeping terms of order up to $\order{\theta^{2}}$, thus retrieving polynomials in $\theta$. The coefficients of the powers of $\theta$ in these polynomials constitute the ordinary differential equations which determine the dependence of all quantities on the radial coordinate, $r$. For a comprehensive description of the formalism that we employed we refer the reader to \cite{loules2024}.

In the following we present a direct comparison between an ideal MHD solution, solution \textbf{I}, and two resistive MHD solutions, \textbf{R1} and \textbf{R2}, which correctly cross the Alfvén critical surface and feature the same boundary conditions at that position. The resistive solutions are ideal ($\eta = 0$) up until a distance $r_{\eta}$ beyond the Alfvén critical surface, located at $r_{A} = 10\, r_{S}$, where the resistivity is switched on. Past $r_{\eta}$, which is $r_{\eta} = 12\, r_{S}$ for \textbf{R1} and $r_{\eta} = 1000\, r_{S}$ for \textbf{R2}, the resistivity increases as a $\tanh$ function of $r - r_{\eta}$, reaching the terminal values of $\eta_{c} = 59\, r_{S}/c$ and $\eta_{c} = 2.4 \times 10^{5}\, r_{S}/c$ in solutions \textbf{R1} and \textbf{R2} respectively. 

\begin{figure*}
\includegraphics[width = 1.0\textwidth]{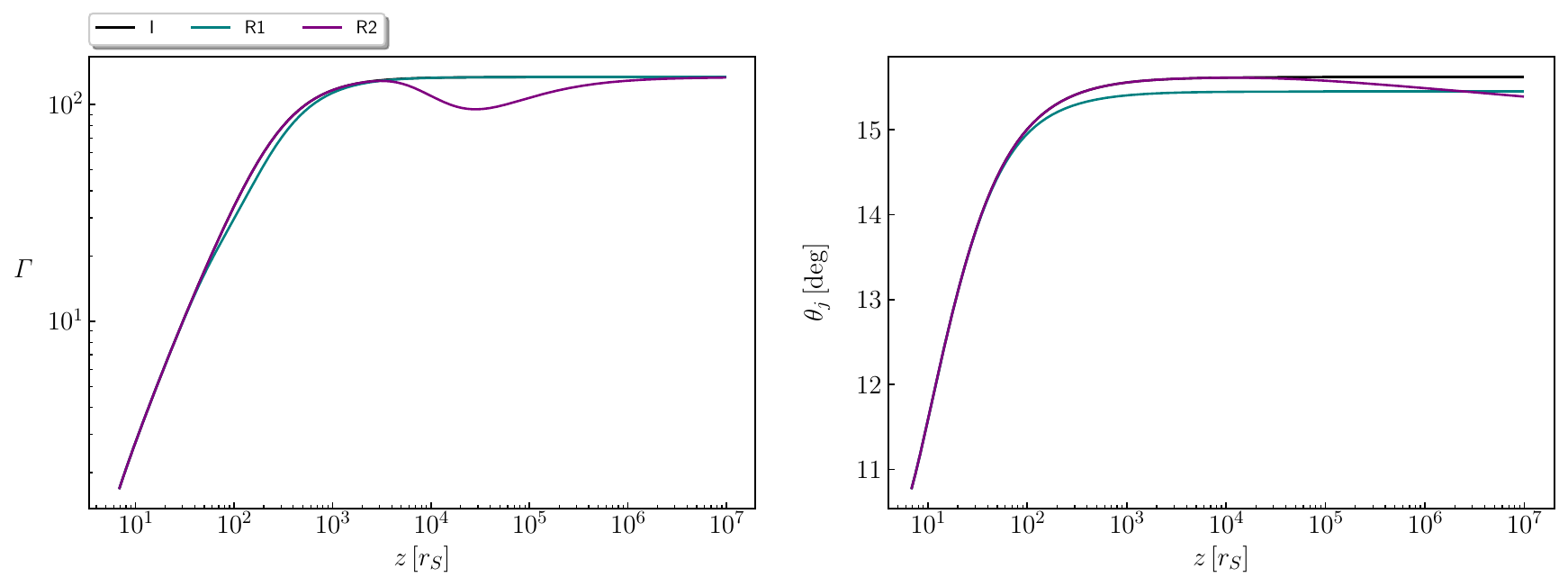}
\caption{Ohmic dissipation affects both the jet's acceleration and collimation profiles. Left panel: jet bulk Lorentz factor for the ideal MHD solution \textbf{I} and resistive MHD solutions \textbf{R1} and \textbf{R2} along the jets' boundary flux lines. Right panel: jet half-opening angle for solutions \textbf{I}, \textbf{R1}, and \textbf{R2}.}
\label{fig:lf}
\end{figure*}

Close to the jet axis of symmetry, the dominant acceleration mechanism is thermal in nature \citep{chantry2018, anastasiadis2024}, with the outflow being accelerated due to the conversion of its thermal internal energy to kinetic. This is the reason why the ideal MHD solution \textbf{I}, which is relativistically hot at its base ($\Theta \sim 10$), reproduces the acceleration profile ($\varGamma \sim z$\footnote{$z \simeq r$ close to the jet axis.}) of the relativistic fireball solution \citep{shemi1993}. As shown in Fig. \ref{fig:lf}, the Lorentz factor profiles of the resistive MHD solutions \textbf{R1} and \textbf{R2} deviate from that of solution \textbf{I} due to the impact of EM dissipation on their thermodynamical properties. Specifically, in the case of solution \textbf{R1}, ohmic dissipation reduces the rate at which the jet's specific enthalpy decreases along the mass flux or magnetic flux lines\footnote{These coincide in ideal MHD jets, as well as in resistive MHD jets close to the jet axis \citep{loules2024}.}, consequently weakening the thermal acceleration of the jet. In solution \textbf{R2}, EM dissipation affects the jet at a much larger distance from its base, where it has already reached its terminal Lorentz factor and its thermal acceleration has ceased ($\xi \simeq 1$). At that distance from the jet base, EM dissipation is strong enough to cause the heating of the outflow along the flux lines, resulting in its deceleration. After the distance at which dissipation becomes negligible, the jet is once again accelerated thermally, until its Lorentz factor becomes equal to the ideal MHD solution's terminal Lorentz factor.

Additionally, the resistive MHD solutions \textbf{R1} and \textbf{R2}, present stronger collimation compared to the ideal MHD solution \textbf{I}, a fact made evident by their half-opening angles, $\theta_{j}$, presented in Fig. \ref{fig:lf}. There are two factors which contribute to the stronger collimation exhibited by the resistive solutions. The first factor is the thermal pressure profile of the resistive solutions. In solution \textbf{R1} the thermal pressure at larger angular distances from the axis decreases at a slower rate compared to the on-axis thermal pressure, with this effect becoming more intense the further away we move from the axis, while in solution \textbf{R2} the off-axis thermal pressure increases over the region of the solution where dissipation is significant. This is due to the profile of the EM dissipation, which follows a $\theta^{2}$ dependence on the polar angle, which makes it stronger at larger angular distances from the axis. As such, there is an extra pressure gradient force perpendicular to the off-axis flux lines which tends to collimate the resistive jets. The second factor is related to the toroidal magnetic field. In the two resistive solutions, the toroidal magnetic field is locally amplified, over the region where dissipation affects the jets, leading to a stronger magnetic collimating force.

\begin{figure}
\includegraphics[width = 0.5\textwidth]{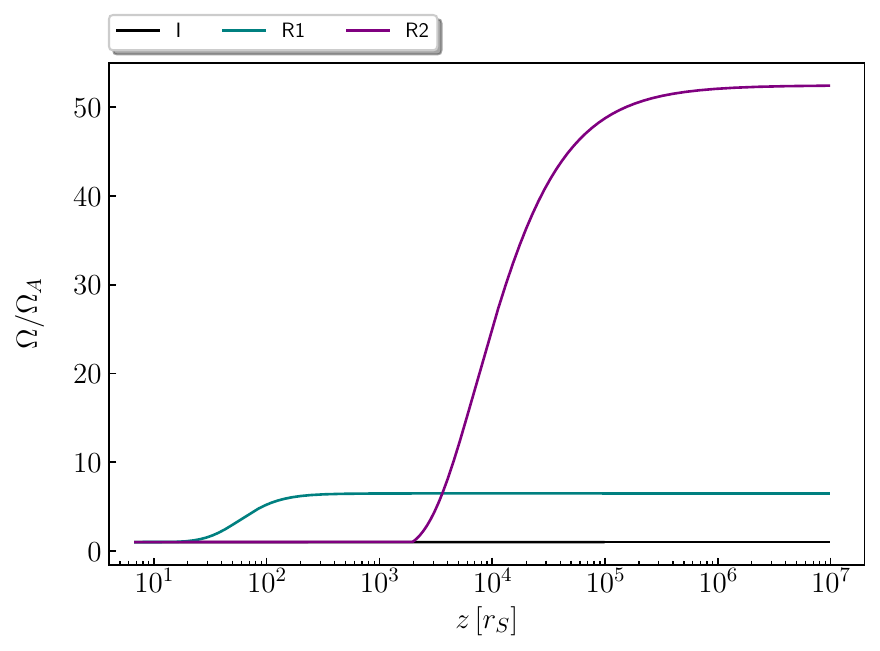}
\caption{Angular velocity of the poloidal magnetic field lines for the ideal MHD solution \textbf{I}, and resistive MHD solutions \textbf{R1}, \textbf{R2}.}
\label{fig:omega}
\end{figure}

The reason behind the emergence of EM dissipation is the increase in the poloidal field line angular velocity $\Omega$ due to the resistivity along the poloidal magnetic field lines, which in steady state, axisymmetric ideal MHD outflows is a field line integral \citep{chantry2018}. $\Omega$ for solutions \textbf{I}, \textbf{R1}, and \textbf{R2} is presented in Fig. \ref{fig:omega}, normalized to its value, $\Omega_{A}$, at the Alfvén critical surface. The increase of $\Omega$ observed in the resistive solutions, leads to a positive electric potential gradient along the off-axis poloidal magnetic field lines, which in ideal MHD jets are lines of constant electric potential \citep{anastasiadis2024}. This gradient of the electric potential leads to the emergence of an electric field component parallel to the poloidal magnetic field lines, which acts as the source of EM dissipation \citep{loules2024}.

\section{Conclusions}

We summarize our findings regarding the effects of EM dissipation on the acceleration and collimation of ultra-relativistic plasma jets as follows:
\begin{itemize}
    \item Resistivity leads to the emergence of ohmic dissipation, which acts as a heat source and affects the rate at which the plasma's thermal energy is converted to kinetic energy along the flux lines, even leading to the heating of the plasma and the deceleration of the jet, as in the case of solution \textbf{R2}. Thus, resistive jets affected by EM dissipation in their acceleration regions (solution \textbf{R1}) do not adhere to the acceleration law predicted by the relativistic fireball model, which is typical for thermally accelerated relativistically hot plasma outflows. On the other hand, when EM dissipation acts at a distance where acceleration has stopped and the jet is coasting at its terminal Lorentz factor, it can cause an increase in the specific enthalpy along the flux lines and as a consequence decelerate the jet (solution \textbf{R2}).

    \item EM dissipation scales with the polar angle as $\theta^{2}$ and as a result affects the thermal pressure of the jet more significantly at larger angular distances from the axis. Outer flux lines experience stronger ohmic heating than the inner ones, resulting in a stronger thermal pressure gradient perpendicular to the flux lines, which tends to enhance the thermal collimation of the resistive jets. Additionally, the toroidal magnetic field of the resistive solutions is amplified over regions of significant dissipation in resistive jets, which strengthens their magnetic collimation.

    \item In resistive jet solutions, the poloidal field line angular velocity, $\Omega$, displays an increase at distances $r > r_{\eta}$, where the resistivity is greater than zero. The increase in $\Omega$ causes the emergence of an electric potential gradient, or equivalently of an electric field component, along the poloidal magnetic field lines. This electric field component is the source of EM dissipation in our resistive jet solutions.
    
\end{itemize}

\begin{acknowledgements}
This work was supported in full by the State Scholarships Foundation (IKY)  scholarship program from the proceeds of the “Nic. D. Chrysovergis” bequest.
\end{acknowledgements}

\bibliographystyle{aa}
\bibliography{bibliography}

\end{document}